\pdfoutput=1

\documentclass{jgaa-art}

\usepackage{amsfonts}
\usepackage{amssymb}
\usepackage{graphicx}
\usepackage{cite}
\usepackage{hyperref}

\newcommand{\Z}{\mathbb{Z}}

\usepackage{fancyhdr}
\begin{document}
\Issue{15}{2}{269}{293}{2011} 
\HeadingAuthor{Eppstein} 
\HeadingTitle{Recognizing Partial Cubes in Quadratic Time} 
\title{Recognizing Partial Cubes in Quadratic Time}

\author[first]{David Eppstein}{eppstein@uci.edu}
\affiliation[first]{Computer Science Department, University of California, Irvine}

\submitted{February 2011}%
\reviewed{May 2011}%
\accepted{May 2011}%
\final{May 2011}%
\published{July 2011}%
\type{Regular paper}%
\editor{Giuseppe Liotta}%

\maketitle

\thispagestyle{empty}

\begin{abstract}
We show how to test whether a graph with $n$ vertices and $m$ edges is a partial cube, and if so how to find a distance-preserving embedding of the graph into a hypercube, in the near-optimal time bound $O(n^2)$, improving previous $O(nm)$-time solutions.
\end{abstract}

\clearpage

\fancyhead[LE]{\thepage~~~Eppstein~~~\it Recognizing Partial Cubes in Quadratic Time}
\fancyhead[RO]{{\sffamily JGAA}, 15(2) 269--293 (2011)~~~\thepage}

\section{Introduction}

A \emph{partial cube} is an undirected and unweighted graph that admits a simple distance-labeling scheme: one can label its vertices by bitvectors in such a way that the distance between any two vertices equals the Hamming distance between the corresponding labels (Figure~\ref{fig:pcube}). That is, the graph can be \emph{isometrically embedded} into a hypercube.

Graham and Pollak~\cite{GraPol-BSTJ-71} were the first to discuss partial cubes, for an application involving communication networks. Since then, these graphs have been shown to model a large variety of mathematical systems:
\begin{itemize}
\item In computational geometry, the adjacencies between the cells in any hyperplane arrangements (represented as a graph with a vertex per cell and an edge between any two cells that share a facet) forms a partial cube~\cite{EppFalOvc-08,Ovc-JMP-05}. As a second geometric example, the flip graphs of triangulations of certain point sets also form partial cubes, a fact that can be used to compute flip distance efficiently for these triangulations~\cite{Epp-JCG-10}.
\item In order theory, the family of total orders over a finite set (with adjacency defined by transpositions), the family of linear extensions of a finite partially ordered set (again with adjacency defined by transpositions), the family of partial orders of a finite set (with adjacency defined by inclusion or removal of an order relation between a single pair of items), and the family of strict weak orders on a finite set (with adjacency defined by inclusion or removal of a separation of the items into two subsets, one of which is less than the other in the weak order) all form partial cubes~\cite{EppFalOvc-08}. For instance, the permutohedron shown in Figure~\ref{fig:pcube} can be interpreted as the graph of total orders of a four-element set.

\begin{figure*}[t]
\centering\includegraphics[width=3.5in]{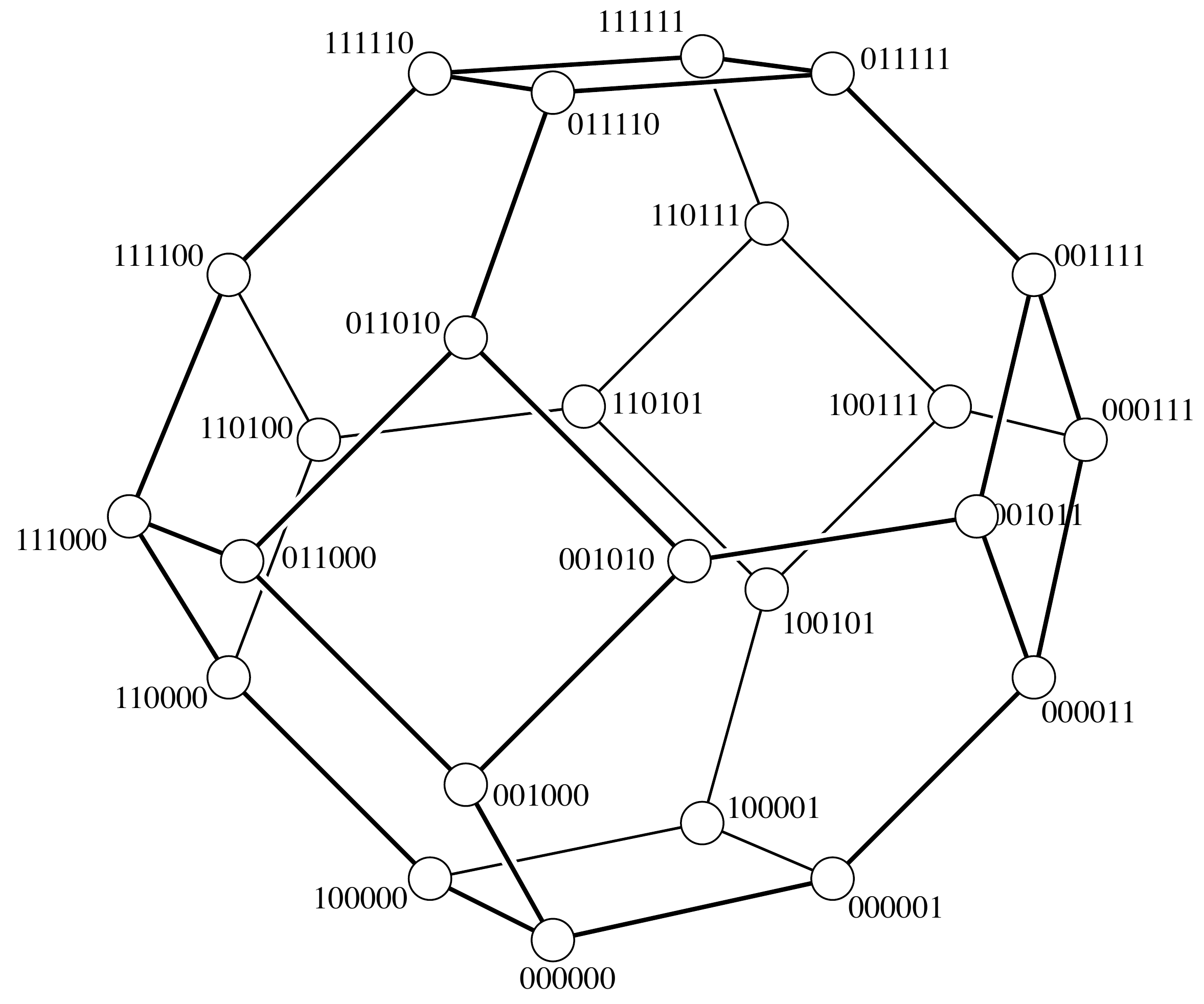}
\caption{A partial cube, with labeled vertices. The distance between any pair of vertices equals the Hamming distance between the corresponding labels, a defining property of partial cubes.}
\label{fig:pcube}
\end{figure*}

\item In the combinatorial study of human learning, antimatroids (called in this context ``learning spaces'') form a standard model of the sets of concepts that a student could feasibly have learned: they are defined by the axioms that such a set may be learned a single concept at a time, and that the union of two feasible sets is another feasible set. In this context, the state space of a learner (a graph with a vertex for each feasible set and an edge connecting any two sets that differ in a single concept) forms a partial cube~\cite{DoiFal-99,EppFalOvc-08}.
\item In organic chemistry, the carbon backbones of certain benzenoid molecules form partial cubes~\cite{PolRou-ICMC-76}, and partial cube labelings of these graphs can be applied in the calculation of their Wiener indices~\cite{KlaGutMoh-JoCIaCS-95}.
\end{itemize}

Partial cubes admit more efficient algorithms than arbitrary graphs for several important problems including unweighted all-pairs shortest paths~\cite{cs.DS/0206033}, and are the basis for several graph drawing algorithms~\cite{Epp-GD-04,Epp-GD-08,Epp-JGAA-08,EppWor-09}.

\subsection{New Results}
In this paper we study the problem of recognizing partial cubes and assigning labels to their vertices. We show that both problems can be solved in time $O(n^2)$, where $n$ is the number of vertices in the input graph. Our algorithm has two phases:
\begin{itemize}
\item In the first phase, we assign bitvector labels to each vertex. It would be straightforward, based on previously known characterizations of partial cubes, to assign a single coordinate of each of these labels by performing a single breadth-first search of the graph; however, the labels may require as many as $n-1$ coordinates, and performing $n-1$ breadth-first searches would be too slow. To speed this approach up, we use the bit-level parallelism inherent in computer arithmetic to assign multiple coordinate values in a single breadth-first pass over the graph. This part of our algorithm depends on a RAM model of computation in which integers of at least $\log n$ bits may be stored in a single machine word, and in which addition, bitwise Boolean operations, comparisons, and table lookups can be performed on $\log n$-bit integers in constant time per operation. The constant-time assumption is standard in the analysis of algorithms, and any machine model that is capable of storing an address large enough to address the input to our problem necessarily has machine words with at least $\log n$ bits.
\item In the second phase, we verify that the labeling we have constructed is indeed distance-preserving. The labels produced in the first phase can be guaranteed to have a Hamming distance that is either equal to the graph distance, or an \emph{underestimate} of the graph distance; therefore, in order to verify that the labeling is distance-preserving, it suffices to construct paths between each pair of vertices that are as short as the Hamming distance between their labels.
To find these paths, we modify an algorithm from previous work with the author and Falmagne~\cite{cs.DS/0206033} that computes all pairs shortest paths in unweighted partial cubes. The modified algorithm either produces paths that are as short as the Hamming distance for each pair of vertices, verifying that the distance labeling is correct, or it detects an inconsistency and reports that the input graph is not a partial cube.
\end{itemize}

Our running time, $O(n^2)$, is in some sense close to optimal, as the output of the algorithm, a partial cube labeling of the input graph, may consist of $\Omega(n^2)$ bits. For instance, labeling a tree as a partial cube requires $n-1$ bits per label. However, in our computational model, such a labeling may be represented in $O(n^2/\log n)$ words of storage, so the trivial lower bound on the runtime of our checking algorithm is $\Omega(n^2/\log n)$. Additionally, in the case of partial cubes that have labelings with few bits per label, or other forms of output than an explicit bitvector labeling of the vertices, even faster runtimes are not ruled out. We leave any further improvements to the running time of partial cube recognition as an open problem.

\subsection{Related Work}

\paragraph{Partial Cube Recognition.}
Since the time they were first studied, it has been of interest to recognize and label partial cubes.
Djokovic~\cite{Djo-JCTB-73} and Winkler~\cite{Win-DAM-84} provided mathematical characterizations of partial cubes in terms of certain equivalence relations on the edges; their results can also be used to describe the bitvector labeling of the vertices of a partial cube, and to show that it is essentially unique when it exists. As Imrich and Klav{\v z}ar~\cite{ImrKla-BICA-93} and Aurenhammer and Hagauer~\cite{AurHag-MST-95} showed, these characterizations can be translated directly into algorithms for recognizing partial graphs in time $O(mn)$, where $m$ and $n$ are respectively the number of edges and vertices in the given graph.\footnote{As we discuss later, for partial cubes, $m\le n\log_2 n$;  the time bound claimed in the title of Aurenhammer and Hagauer's paper is $O(n^2\log n)$, which is therefore slower than $O(mn)$, but it is not hard to see that their algorithm actually takes time $O(mn)$.} Since then there has been no improvement to the $O(mn)$ time bound for this problem until our work.

\paragraph{Special Subclasses of Partial Cubes.}
Several important families of graphs are subclasses of the partial cubes, and can be recognized more quickly than arbitrary partial cubes:
\begin{itemize}
\item Every tree is a partial cube~\cite{math.CO/0402246}, and obviously trees can be recognized in linear time.
\item \emph{Squaregraphs} are the planar graphs that can be drawn in the plane in such a way that every bounded face has four sides and every vertex with degree less than four belongs to the unbounded face. Every squaregraph is a partial cube, and squaregraphs may be recognized in linear time~\cite{BanCheEpp-SJDM-10}.
\item A \emph{median graph} is a graph in which, for every three vertices, there is a unique median vertex that belongs to shortest paths between each pair of the three vertices~\cite{Ava-PAMS-61,BirKis-BAMS-47,Neb-CMUC-71}. The graphs of distributive lattices are median graphs~\cite{BirKis-BAMS-47}; median graphs also arise from the solution sets of 2-satisfiability problems~\cite{Fed-MotAMS-95} and the reconstruction of phylogenetic trees~\cite{BanMacRic-MPE-00,Bun-MAHS-71}. Based on earlier work by Hagauer et al.~\cite{HagImrKla-TCS-99}, Imrich et al.~\cite{ImrKlaMul-SJDM-99} showed that the times for median graph recognition and for triangle-free graph recognition are within polylogarithmic factors of each other. Applying the best known algorithm for triangle detection, based on fast matrix multiplication~\cite{AloYusZwi-JACM-95} yields a time bound of $O(n^{1.41})$ for median graph recognition.
\item Bre{\v s}ar et al.~\cite{BreImrKla-DAM-03} discuss several other classes of partial cubes that are closely related to the median graphs and may be recognized in $O(m\log n)$ time.
\end{itemize}
 
\paragraph{Other Distance Labeling Schemes.}
The assignment of bitvectors to vertices in a partial cube is a form of a \emph{distance labeling scheme}, an assignment of labels to vertices in arbitrary graphs that allows distances to be computed from the labels~\cite{GavPelPer-Algs-04}.
Although bitvectors provide a convenient representation of distances in partial cubes, they are not the only possible scheme for distance labeling, and other schemes may be more concise. The \emph{isometric dimension} of a partial cube is the number of bits needed in each bitvector label, and as discussed above it may be as high as $n-1$.

Every partial cube may be embedded in a distance-preserving way into an integer lattice $\Z^d$ of some dimension $d$. One such labeling simply uses each bit of a bitvector labeling as a coordinate in $\Z^d$; however, some graphs may be embeddable into integer lattices of much lower dimension than their isometric dimension. For instance, a path graph can be embedded into $\Z$, and given one-dimensional coordinates that accurately describe the graph distances, despite having an isometric dimension of $n-1$. The \emph{lattice dimension} of a partial cube is the minimum number $d$ for which the graph admits a distance-preserving embedding into  $\Z^d$. The lattice dimension, and an embedding of that dimension, may be found in polynomial time using an algorithm based on graph matching~\cite{Epp-EJC-05}, but this algorithm depends on having as input a bitvector labeling and is slower than the algorithm we describe here, so it does not form the basis of an efficient partial cube recognition algorithm.

It may also be possible to express a partial cube as a distance-preserving subgraph of a Cartesian product of trees, using many fewer trees than the lattice dimension of the graph. For instance, the star $K_{1,n-1}$ has lattice dimension $\lceil\frac{n-1}{2}\rceil$ despite being a single tree~\cite{math.CO/0402246}. Any individual tree admits a distance labeling scheme with $O(\log^2 n)$-bit labels~\cite{GavPelPer-Algs-04}; even more concisely, it is possible to assign $O(\log n)$-bit identifiers to the nodes of a tree in such a way that pairwise distances can be looked up in constant time per query, based on lowest common ancestor data structures~\cite{BenFar-LATIN-00,HarTar-SJC-84}. Therefore, finding small tree product representations would be of interest as a method of efficient distance representation in these graphs. However, although it is possible to find a representation as a subgraph of a product of two trees in linear time, when such a representation exists~\cite{BanCheEpp-10}, it is NP-hard to find optimal representations using larger numbers of trees or even to find accurate approximations of the optimal number of trees needed in such a representation, due to a reduction from graph coloring~\cite{BanVel-PLMS-89}.

\subsection{Organization}
The remainder of this paper is organized as follows. In Section~\ref{sec:characterization} we review a characterization of partial cubes by Winkler~\cite{Win-DAM-84}. Winkler characterizes partial cubes in terms of an equivalence relationship defined on the edges of the graph by an inequality between sums of pairs of distances; this characterization is central to past partial cube recognition algorithms as well as our own. In this section we also review other standard results on partial cubes needed in our work. In Section~\ref{sec:single} we describe how to find a single bit within each vertex label of a partial cube by using Winkler's characterization as part of an algorithm based on breadth-first search, and in Section~\ref{sec:multiple} we show how to find multiple bits of each label by a single pass of breadth-first search. In Section~\ref{sec:all} we show how this method leads to an efficient algorithm for finding the complete bitvector labels of each vertex. In Section~\ref{sec:apsp} we review our previous algorithm for all-pairs shortest paths in partial cubes and examine its behavior on graphs that might not be partial cubes, and in Section~\ref{sec:verify} we show how to use this algorithm to test whether the labeling we have constructed is valid. Section~\ref{sec:implementation} reports on a proof-of-concept implementation of our algorithms. We conclude in Section~\ref{sec:conclusions}.

\section{Preliminaries}
\label{sec:characterization}

\begin{figure*}[t]
\centering\includegraphics[width=4in]{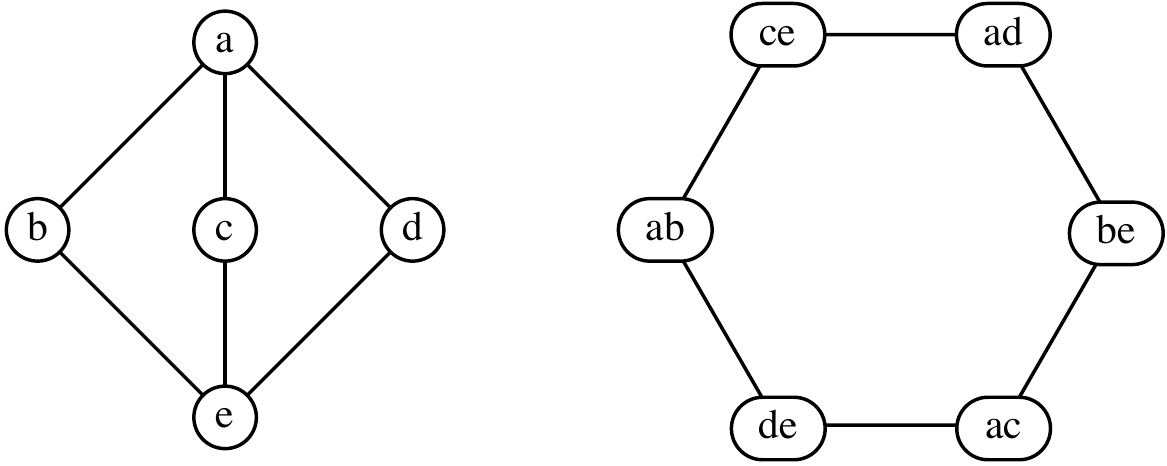}
\caption{An example of Winkler's relationship, for the graph $G=K_{2,3}$ (left). In this graph, each edge is related to the two other edges that it does not share an endpoint with; the right side of the figure shows pairs of edges that are related to each other. In this graph, $\sim_G$ is not an equivalence relationship; for instance, $ab \sim_G ce \sim_G ad$, but $ab\not\sim_G ad$. Thus, by Winkler's characterization, $K_{2,3}$ is not a partial cube.}
\label{fig:Winkler-K23}
\end{figure*}

The characterizations of partial cubes by Djokovic~\cite{Djo-JCTB-73} and Winkler~\cite{Win-DAM-84} both depend on defining certain relations on the edges of the graph that, in the case of partial cubes, can be shown to be equivalence relations. Moreover, although Djokovic's and Winkler's relations may differ from each other on arbitrary graphs, they are identical on partial cubes. It will be more convenient for our purposes to start with the formulation of Winkler. Therefore, following Winkler, define a relation $\sim_G$ on the edges of an undirected graph $G$, by setting $pq \sim_G rs$ if and only if $d(p,r)+d(q,s)\ne d(p,s)+d(q,r)$, where $d$ denotes the number of edges in the shortest path between two vertices.

This relation is automatically reflexive in any graph without self-loops: for every edge $pq$, $pq\sim_G pq$. It is also symmetric: if $pq\sim_G rs$ then $rs\sim_G pq$, and vice versa.
It also does not depend on the ordering of the two endpoints of the edges it relates. These are two of the three defining properties of an equivalence relation, the third being transitivity.

For example, if $pqrs$ form a path, with no additional edges connecting these four vertices, then $pq\not\sim_G rs$ because $d(p,r)+d(q,s)=2+2=3+1=d(p,s)+d(q,r)$. On the other hand, if $pqrs$ form a 4-cycle, again with no additional edges, then $pq\sim_G rs$ because $d(p,r)+d(q,s)=2+2\ne 1+1=d(p,s)+d(q,r)$. Figure~\ref{fig:Winkler-K23} shows a more complicated example of a graph $K_{2,3}$ with six edges, and the Winkler relation among these edges.

\begin{lemma}[Winkler]
\label{lem:winkler}
Graph $G$ is a partial cube if and only if $G$ is bipartite and $\sim_G$ is an equivalence relation.
\end{lemma}

Referring again to the example in Figure~\ref{fig:Winkler-K23}, the transitive property does not hold: for instance, $ab\sim_G ce$, and $ce\sim_G ad$, but $ab\not\sim_G ad$. Therefore, for this example, $\sim_G$ is not an equivalence relation and Winkler's lemma tells us that the graph $K_{2,3}$ shown in the figure is not a partial cube.

We will use $[e]$ to denote the set of edges related to an edge $e$ by $\sim_G$ (that is, in the case that $G$ is a partial cube, the equivalence class of $e$).

If $G$ is a partial cube, and $e=pq$ is any edge of $G$, then
let $S_{pq}$ denote the set of vertices nearer to $p$ than to $q$, and $S_{qp}$ denote the set of vertices nearer to $q$ than to $p$. (There can be no ties in a bipartite graph.) The sets $S_{pq}$ and $S_{qp}$ were called \emph{semicubes} in our algorithm for lattice embeddings of partial cubes~\cite{Epp-EJC-05}, where they play a key role, and they are also central to Djokovic's and Winkler's characterizations of partial cubes. Equivalently, $S_{pq}$ must consist of the vertices whose labels match that of $p$ in the coordinate at which the labels of $p$ and $q$ differ, and $S_{qp}$ must consist of the vertices whose labels match that of $q$ in the same coordinate. The edges separating these two subsets are exactly the edges in $[e]$, and both of these two subsets must be connected (since every pair of vertices in one of these two subsets can be connected by a path that does not change the label at the coordinate that they share with $p$ or~$q$).

Thus, as shown by Winkler, in a partial cube, each equivalence class $[e]$ forms an edge cut partitioning the graph into two connected components, and the partial cube labeling for $G$ has a coordinate $i$ such that the $i$th bit in all labels for vertices in one of the two components is 0, and the same bit in all labels for vertices in the other component is 1. The dimension of the partial cube labeling (the isometric dimension of the graph) equals the number of equivalence classes of $\sim_G$, and the labeling itself is essentially unique up to symmetries of the hypercube.

It will be important for our algorithms to observe that any partial cube with $n$ vertices has at most $n\log n$ edges. This appears to be folklore (see e.g. Lemma~3 of Matou\v{s}ek~\cite{Mat-Comb-06}) but we repeat for completeness a proof, copied (in different terminology) from Lemma~4 of~\cite{cs.DS/0206033}.

\begin{lemma}
In any $n$-vertex partial cube, the number of edges is at most $n\log_2 n$.
\end{lemma}

\begin{proof}
We apply induction on the isometric dimension.
As a base case, if there is only one vertex there can be no edges.
Otherwise, let $e=uv$ be any edge in the graph, partition the graph into two components $G_u$ and $G_v$, and assume without loss of generality that $|G_u|\le |G_v|$. Then both $G_u$ and $G_v$ induce partial cubes, which have a number of edges that can be bounded by induction to the same formula of their numbers of vertices. In addition, the number of edges in $[e]$ is at most $|G_u|$, because each edge has an endpoint in $G_u$ and each vertex in $G_u$ can be the endpoint for at most one edge. (If it were the endpoint of two edges in $[e]$, the other endpoints of those edges would have equal labels, contradicting their nonzero distance from each other.)

So, if $M(n)$ denotes the maximum number of edges in any $n$-vertex partial cube, we have a recurrence
$$M(n)\le\max \bigl\{ M(a)+M(b)+\min(a,b) \mid a+b=n \bigr\}$$
which can be used in an induction proof to derive the desired bound.
\end{proof}

\section{Finding a single edge class}
\label{sec:single}

Given a graph $G$ and an edge $pq$ of $G$, it is straightforward to construct the set $[pq]$ of edges related to $pq$ by $\sim_G$: perform two breadth first searches, one starting from $p$ and another starting from $q$, using the resulting breadth first search trees to calculate all distances from $p$ or $q$ to other vertices of the graph, and then apply the definition of Winkler's relation $\sim_G$ to test whether each other edge of the graph belongs to $[pq]$ in constant time per edge. We begin the description of our algorithm by showing how to simplify this construction: we may find $[pq]$ by an algorithm that performs only a single breadth first search rather than two searches. Moreover, we need not calculate any distances as part of this computation. This simplification will be an important step of our overall result, as it will eventually allow us to construct multiple equivalence classes of edges simultaneously, in less time than it would take to perform each construction separately.

Our technique is based on the following observation:

\begin{lemma}
Let $pq$ be an edge in a bipartite graph $G$. Then $pq \sim_G rs$ if and only if exactly one of $r$ and $s$ has a shortest path to $p$ that passes through $q$.
\end{lemma}

\begin{proof}
If neither $r$ nor $s$ has such a path, then $d(q,r)=d(p,r)+1$ and $d(q,s)=d(p,s)+1$, so $d(p,r)+d(q,s)=d(p,r)+1+d(p,s)=d(q,r)+d(p,s)$ by associativity of addition, and $pq\not\sim_G rs$. Similarly, if both $r$ and $s$ have such paths, then $d(q,r)=d(p,r)-1$ and $d(q,s)=d(p,s)-1$, so $d(p,r)+d(q,s)=d(p,r)-1+d(p,s)=d(q,r)+d(p,s)$. Thus in neither of these cases can $pq$ and $rs$ be related.
If, on the other hand, exactly one of $r$ and $s$ has such a path, we may assume (by swapping $r$ and $s$ if necessarily that it is $r$ that has the path through $q$. Then $d(q,r)=d(p,r)-1$ while $d(q,s)=d(p,s)+1$, so $d(p,r)+d(q,s)=d(p,r)+d(p,s)+1\ne d(p,r)-1+d(p,s)=d(q,r)+d(p,s)$, so in this case $pq\sim_G rs$.
\end{proof}

Thus, to find the edge class $[pq]$ in a bipartite graph $G$, we may perform a breadth first search rooted at $p$, maintaining an extra bit of information for each vertex $v$ traversed by the search: whether $v$ has a shortest path to $p$ that passes through $q$. This bit is set to false initially for all vertices except for $q$, for which it is true. Then, when the breadth first search traverses an edge from a vertex $v$ to a vertex $w$, such that $w$ has not yet been visited by the search (and is therefore farther from $p$ than $v$), we set the bit for $w$ to be the disjunction of its old value with the bit for~$v$. Note that we perform this update for all edges of the graph, regardless of whether the edges belong to any particular breadth first search tree.

Recall that $S_{pq}$ denotes the set of vertices nearer to $p$ than to $q$. It will be important to the correctness of our algorithm to make the following additional observation. 

\begin{lemma}
\label{lem:class-cut}
If $G$ is bipartite, then for any edge $pq$ the semicubes $S_{pq}$ and $S_{qp}$ partition $G$ into two subsets, and the edge class $[pq]$ forms the cut between these two semicubes.
\end{lemma}

\begin{proof}
This follows immediately from the previous lemma, since $S_{qp}$ consists exactly of the vertices that have a shortest path to $p$ passing through $q$.
\end{proof}

We remark that this description of edge classes $[pq]$ in terms of semicubes  is very close to Djokovic's original definition of an equivalence relation on the edges of a partial cube. Thus, for bipartite graphs, Winkler's definition (which we are following here) and Djokovic's definition can be shown to coincide.

\section{Finding several edge classes}
\label{sec:multiple}

As we now show, we can apply the technique described in the previous section to find several edge classes at once. Specifically, we will find classes $[pq]$ for each neighbor $q$ of a single vertex $p$, by performing a single breadth first search rooted at $p$.

\begin{lemma}
Let $pq$ and $pr$ be edges in a bipartite graph $G$. Then $pq\not\sim_G pr$.
\end{lemma}

\begin{proof}
By bipartiteness, $d(q,r)=2$, so $d(p,p)+d(q,r)=2=1+1=d(p,r)+d(q,p)$.
\end{proof}

Our algorithm will need efficient data structures for storing and manipulating bit vectors, which we now describe. As described in the introduction, we assume throughout that arithmetic and bitwise Boolean operations on integers of at least $\log n$ bits, as well as array indexing operations, are possible in constant time.

\begin{lemma}
\label{lem:bv}
Let $k$ be a given number, and let $K=1+k/\log n$.
Then it is possible to store bitvectors with $k$ bits each in space $O(K)$ per bitvector, and perform disjunction operations and symmetric difference operations in time $O(K)$ per operation. In addition, in time $O(K)$ we can determine whether a bitvector contains any nonzero bits. If it does, in time $O(K)$ we can determine whether it has exactly one nonzero bit, and if so find the index of that bit, using a single precomputed external table of size $n$.
\end{lemma}

\begin{proof}
We store a bitvector in $\lceil K\rceil$ words, by packing $\log n$ bits per machine word.
Disjunction and symmetric difference can be performed independently on each of these words.
To test whether a bitvector is nonzero, we use a comparison operation to test whether each of its words is nonzero. To test whether a bitvector has exactly one nonzero bit, and if so find out which bit it is, we again use comparisons to test whether there is exactly one word in its representation that is nonzero, and then look up that word in a table that stores either the index of the nonzero bit (if there is only one) or a flag value denoting that there is more than one nonzero bit.
\end{proof}

\begin{figure*}[t]
\centering\includegraphics[scale=0.4]{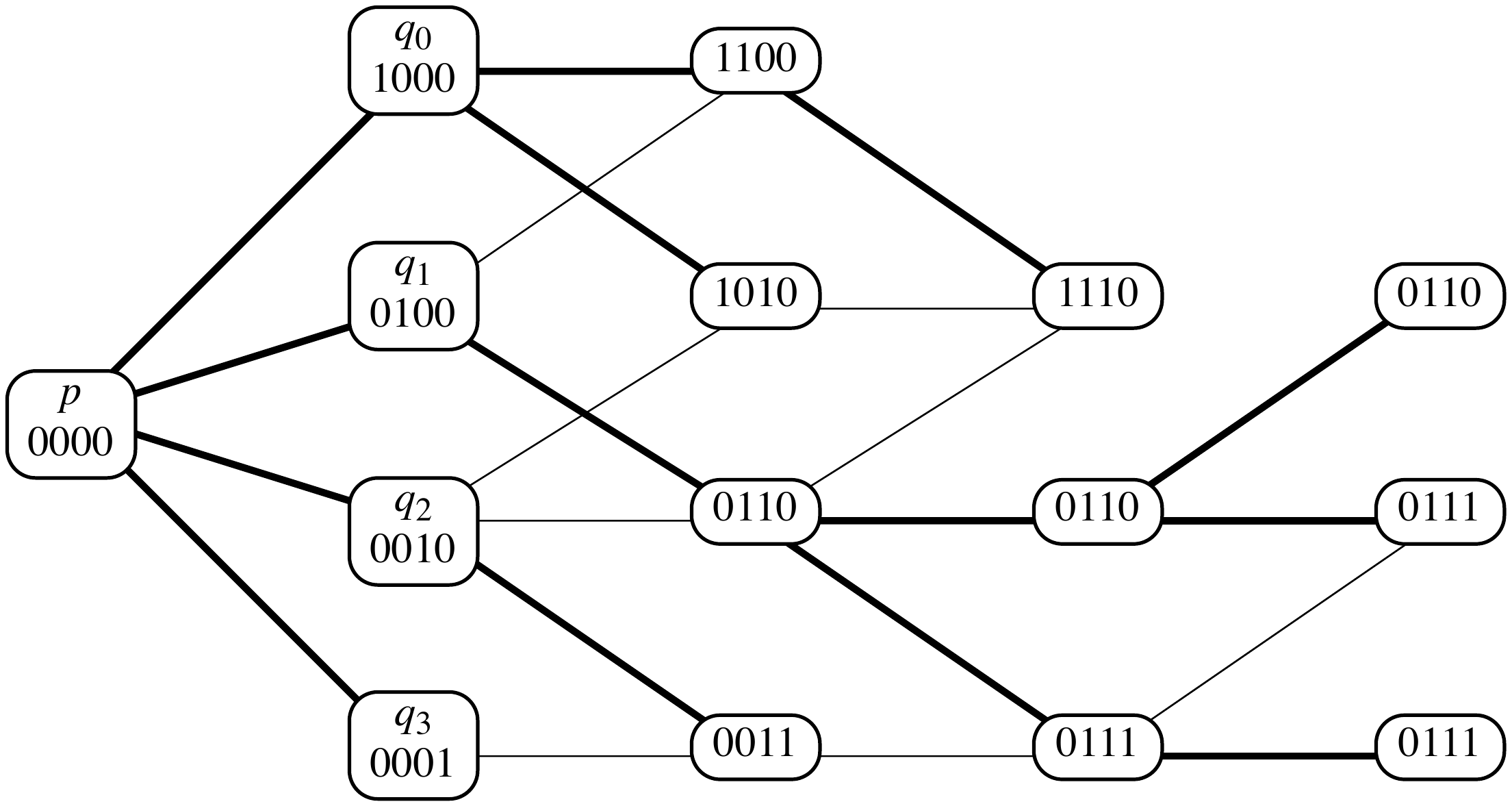}
\caption{The vertex-labeling stage of the algorithm of Lemma~\ref{lem:some-classes}. The breadth first search tree edges are shown darker than the other edges; the left-to-right placement of the vertices is determined by their distance from the starting vertex $p$. Except for the neighbors $q_i$ of the starting vertex, the bitvector shown for each vertex is the disjunction of the bitvectors of its neighbors to the left.}
\label{fig:bfs-bits}
\end{figure*}

We are ready to specify the main algorithm of this section, for finding a collection of edge classes of our supposed partial cube.

\begin{figure*}[t]
\centering\includegraphics[scale=0.4]{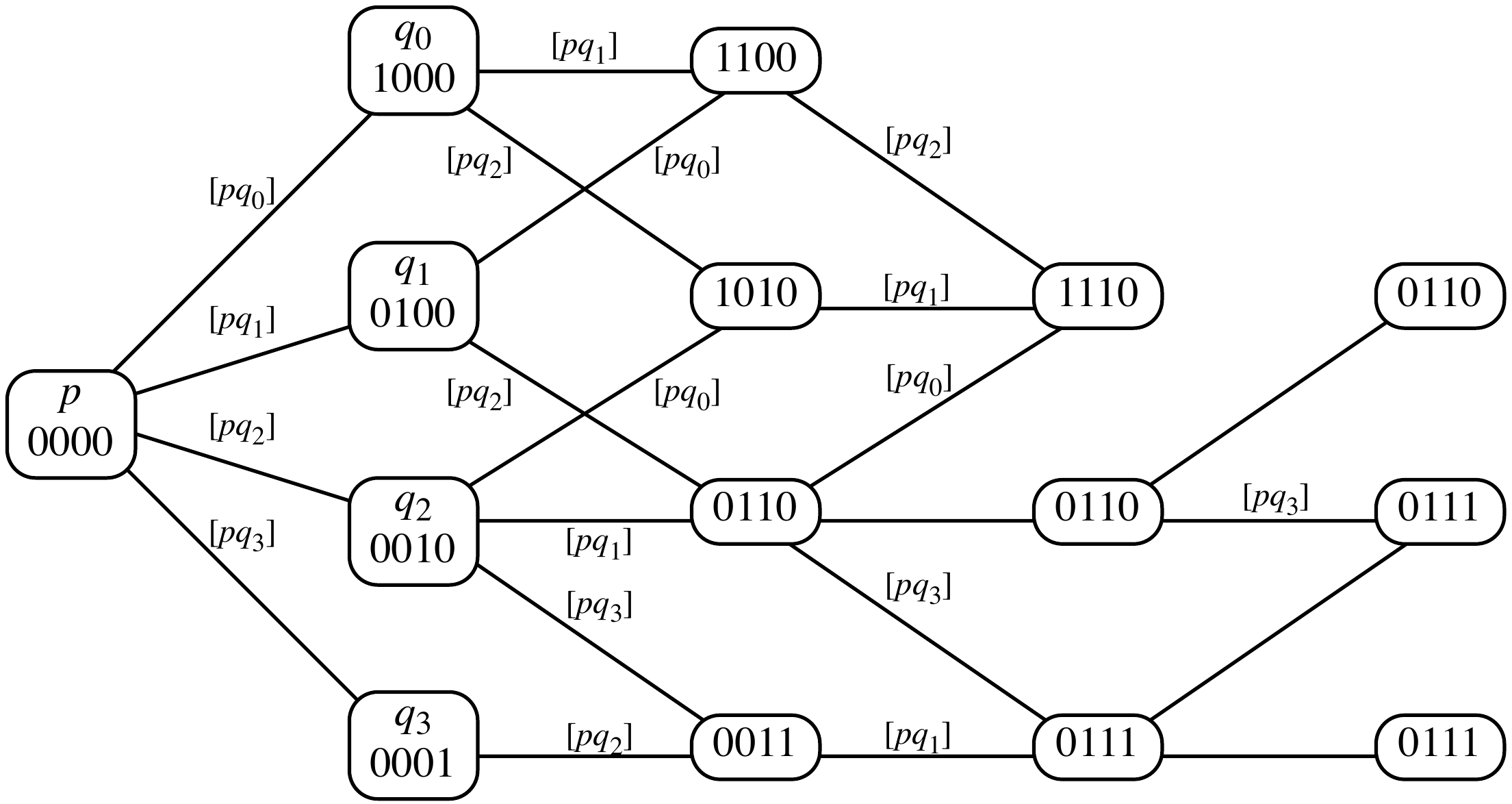}
\caption{The edge-labeling stage of the algorithm of Lemma~\ref{lem:some-classes}. If the bitvectors of the endpoints of an edge differ only in their $i$th bits, the edge is included in class $[pq_i]$. If the bitvectors of the endpoints are the same, the edge is not included in any class. If there were an edge that had bitvectors differing in more than one bit, the graph would not be a partial cube.}
\label{fig:bfs-edges}
\end{figure*}

\begin{lemma}
\label{lem:some-classes}
Let $G$ be any graph with $n$ vertices and $m$ edges. Then there is an algorithm which either determines that $G$ is not a partial cube (taking time at most $O(n^2)$ to do so) or finds a collection ${\mathcal E}$ of disjoint sets of edges $[e_i]$, with $|E|\ge 2m/n$, taking time $O(|{\mathcal E}|\cdot n)$ to do so where $|{\mathcal E}|$ is the number of sets in the collection. In the latter case, the algorithm can also label each vertex of $G$ by the set of semicubes it belongs to among the semicubes corresponding to the edges $e_i$, in the same total time.
\end{lemma}

\begin{proof}
We first check that $G$ is bipartite; if not, it cannot be a partial cube. We also check that its number of edges is at most $n\log_2 n$, and if not we again report that it is not a partial cube. We then let $p$ be a vertex of maximum degree in $G$. We denote by $d$ the degree of $p$, which must be at least $2m/n$. We denote the $d$ neighbors of $p$ in $G$ by $q_i$, for an index $i$ satisfying $0\le i<d$.

We create, for each vertex of $G$, a data structure $D_v$ with $d$ bits $D_v[i]$.
Bit $D_v[i]$ will eventually be 1 if $v$ has a shortest path to $p$ that passes through $q_i$ (that is, if $v\in S_{q_ip}$); initially, we set all of these bits to 0 except that we set $D_{q_i}[i]=1$.
Next, we perform a breadth first traversal of $G$, starting at $p$. When this traversal finds an edge from a vertex $v$ to a vertex $w$ that has not yet been traversed (so $w$ is farther from $p$ than $v$), it sets all bits $D_w[i]$ to be the disjunction of their previous values with the corresponding bits $D_v[i]$, as shown in Figure~\ref{fig:bfs-bits}.

Finally, once the breadth first search is complete and all data structures $D_v$ have reached their final values, we examine each edge $vw$ in the graph. If $D_v=D_w$, we ignore edge $vw$, as it will not be part of our output collection. Otherwise, we compute a bitvector $B$ as the symmetric difference of $D_v$ and $D_w$. If $B$ contains two or more nonzero bits $B[i]$ and $B[j]$, then $vw$ belongs to both $[pq_i]$ and $[pq_j]$, and $G$ cannot be a partial cube; if we ever encounter this condition we terminate the algorithm and report that the graph is not a partial cube. Otherwise, we assign $vw$ to the class $[pq_i]$ for which $B[i]$ is nonzero. Figure~\ref{fig:bfs-edges} shows this assignment of edges to classes for the example graph shown in Figure~\ref{fig:bfs-bits}.

The result of this algorithm is a collection ${\mathcal E}$ of disjoint sets of edges $[pq_i]$, as the lemma requires; the number of sets in the collection is $d$. All stages of the algorithm perform $O(m)$ steps, each one of which involves at most $O(1)$ of the bitvector operations described by Lemma~\ref{lem:bv}, so the total time is $O(m(1 + d/\log n))=O(d(m/d+m/\log n))=O(dn)$. Since $d\le n$, this bound is $O(n^2)$, as the lemma states for the time taken when the input is determined not to be a partial cube, and since $d=|{\mathcal E}|$ the time is $O(|{\mathcal E}|n)$ when the algorithm successfully constructs a set of edge classes.

The semicube labeling output described by the statement of the lemma is represented by the data structures $D_v$ computed as part of the algorithm.
\end{proof}

\section{Finding all edge classes}
\label{sec:all}

In order to recognize a partial cube, we need to partition its edges into equivalence classes of the relation $\sim_G$, and then verify that the resulting labeling is correct. The algorithm of the previous section allows us to find some of these equivalence classes efficiently, but as it depends for its efficiency on starting from a high degree vertex we will not necessarily be able to use it multiple times on the same graph. In order to reapply the algorithm and find all equivalence classes efficiently, as we now describe, we will need to remove from the graph the parts we have already recognized.

\begin{lemma}
\label{lem:contraction-relation}
Let $G$ be a partial cube, let $pq$ be an edge in $G$, and let $G'$ be the graph formed from $G$ by contracting all edges in $[pq]$. For any edges $e$ and $f$ in $G$, neither of which belong to $[pq]$, let $e'$ and $f'$ denote the corresponding edges in $G'$. Then $e\sim_G f$ if and only if $e'\sim_{G'} f'$.
\end{lemma}

\begin{proof}
If $e$ and $f$ are not in $[pq]$, by Lemma~\ref{lem:class-cut}, either both edges connect vertices in one of the two semicubes $S_{pq}$ and $S_{qp}$, or one edge is entirely in one semicube and the other edge is in the other semicube. If both are in the same semicube, then no shortest path from any vertex of $e$ to any vertex of $f$ can use an edge of $[pq]$ (for if it did, that crossing would increase rather than decrease the Hamming distance of the path vertex's labels), so the distances $d(x,y)$ used in the definition of $\sim_{G'}$ remain unchanged from those used to define $\sim_G$. If, on the other hand, $e$ and $f$ are in opposite semicubes, then by similar reasoning every shortest path from an endpoint of $e$ to a vertex of $f$ must use exactly one edge of $[pq]$, and each distance $d(x,y)$ used in the definition of $\sim_{G'}$ is exactly one smaller than the corresponding distance in the definition of $\sim_G$. Since we are subtracting two units of distance total from each side of the inequality by which $\sim_{G'}$ is defined, it remains unchanged from $\sim_G$.
\end{proof}

\begin{lemma}
\label{lem:contraction-pcube}
Let $G$ be a partial cube, let $pq$ be an edge in $G$, and let $G'$ be the graph formed from $G$ by contracting all edges in $[pq]$.
Then $G'$ is a partial cube, the equivalence classes of edges in $G'$ correspond with those in $G$ except for $[pq]$, and the vertex labeling of $G'$ is formed by omitting the coordinate corresponding to $[pq]$ from the vertex labeling of $G$.
\end{lemma}

\begin{proof}
By Lemma~\ref{lem:contraction-relation}, $\sim_{G'}$ coincides with $\sim_G$ on the remaining edges; thus, it is an equivalence relation, $G'$ is a partial cube, and its equivalence classes correspond with those of $G$. Since the vertex labeling is formed from the semicubes of $G'$, which are derived from the cuts formed by equivalence classes of edges, they also correspond in the same way.
\end{proof}

\begin{lemma}
Any partial cube with $n$ vertices has at most $n-1$ edge equivalence classes.
\end{lemma}

\begin{proof}
Choose arbitrarily a vertex $v$. For any edge equivalence class $[pq]$, with $p$ closer to $v$ than $q$ is, any shortest path from $v$ to $q$ must pass through an edge in $[pq]$ by Lemma~\ref{lem:class-cut}. In particular, if $T$ is a breadth-first spanning tree of the graph, rooted at $v$, $T$ must include an edge in $[pq]$. But $T$ has only $n-1$ edges, and each equivalence class is represented by at least one edge in $T$, so there can be at most $n-1$ equivalence classes.
\end{proof}

\begin{figure*}[t]
\centering\includegraphics[scale=0.4]{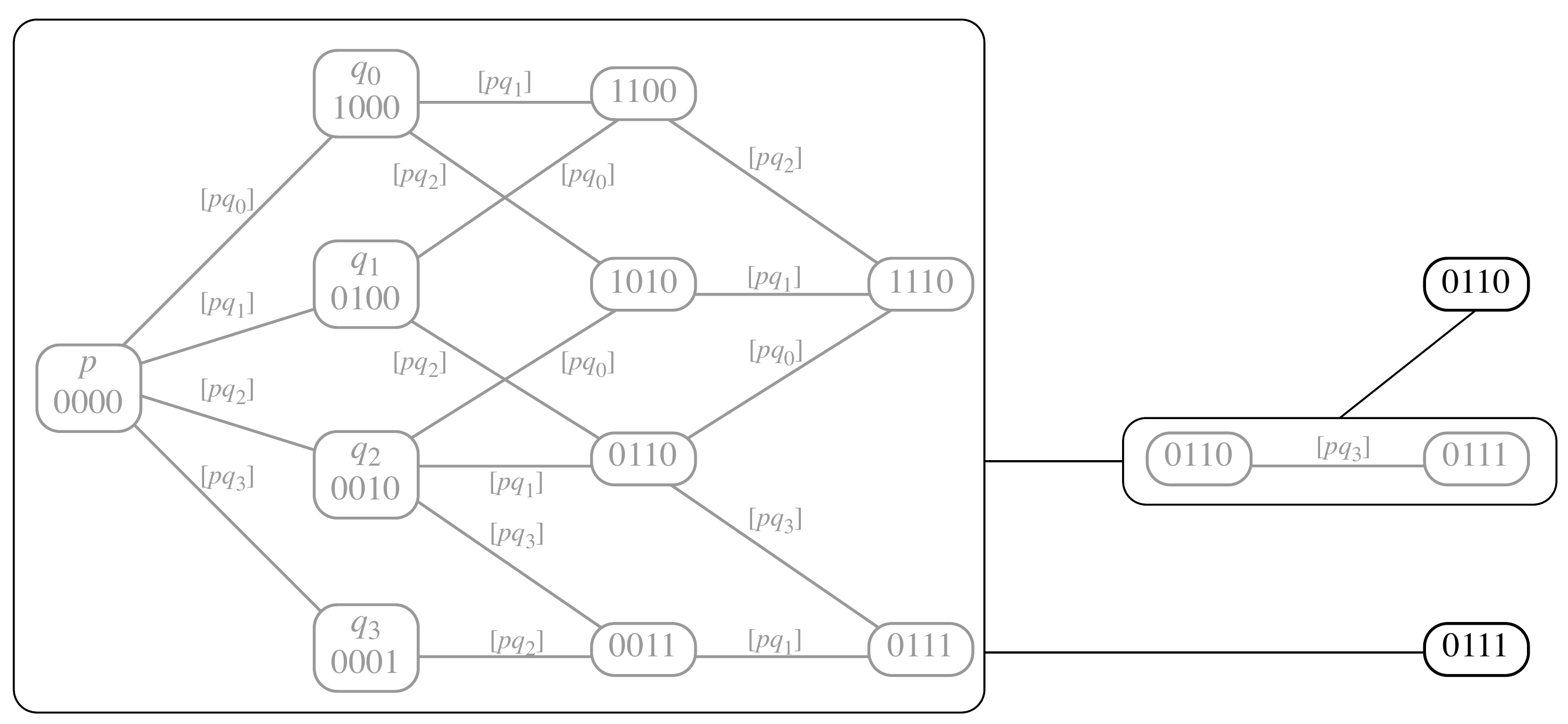}
\caption{A four-vertex path graph formed by contracting the labeled edges from Figure~\ref{fig:bfs-edges}.}
\label{fig:bfs-contracted}
\end{figure*}

Our algorithm for partitioning the edges of a graph $G$ into classes (that, if $G$ is a partial cube, will be the equivalence classes of $\sim_G$) and simultaneously labeling the vertices of $G$ with bitvectors (that, if $G$ is a partial cube, will be a correct partial cube labeling for $G$) performs the following steps. As part of the algorithm, we set a limit $L$ on the number of equivalence classes it can output; for our initial call to the algorithm, we set $L=n-1$, but it will be smaller in the recursive calls the algorithm makes to itself.
\begin{itemize}
\item If $G$ has one vertex and no edge, we report that it is a partial cube, label its vertex with a bitvector of length zero, and return an empty set of edge equivalence classes.
\item We find the maximum degree $d$ of a vertex in $G$ and test whether $d$ exceeds the remaining limit on the number of allowed equivalence classes. If it does, we terminate the algorithm and report that $G$ is not a partial cube.
\item We apply the algorithm of Lemma~\ref{lem:some-classes} to find a set ${\mathcal E}$ of $d$ edge classes of $G$. If this algorithm terminates and reports that $G$ is not a partial cube, we do likewise.
\item We contract all edges that belong to classes in ${\mathcal E}$, and remove any self-loops or multiple adjacencies in the resulting contracted graph. As we do so, we maintain a correspondence of edges in $G$ with the edges representing them in the contracted graph $G'$, and between vertices in $G$ and the corresponding vertices in $G'$. If a set of edges in $G$ corresponds to a multiple adjacency in $G'$, we represent them all by the same single edge in $G'$. If an edge in $G$ corresponds to a self-loop in $G'$, and does not belong to one of the classes in ${\mathcal E}$, we terminate the algorithm and report that $G$ is not a partial cube. Figure~\ref{fig:bfs-contracted} shows the smaller contracted graph $G'$ resulting from this step of the algorithm.
\item We apply the same algorithm recursively, to partition the edges and label the vertices of $G'$. In this recursive call we limit the algorithm to output at most $L-d$ equivalence classes. If this algorithm terminates and reports that $G'$ is not a partial cube, we terminate and report that $G$ is also not a partial cube.
\item We propagate the labels and partition of $G'$ back to the vertices and edges of $G$, using the correspondence created when we contracted $G$ to form $G'$.
\item To form the list of equivalence classes of edges for $G$, we concatenate the list of equivalence classes for $G'$ (with the edges replaced by the edges they correspond to in $G$) with the separate list of classes ${\mathcal E}$.
\item To form the vertex label for each vertex $v$ of $G$, we concatenate the bitvector for the vertex corresponding to $v$ in $G'$ with the bitvector $D_v$ found by the algorithm of Lemma~\ref{lem:some-classes}.
\end{itemize}

As an example, if we apply our algorithm to the graph of Figures~\ref{fig:bfs-bits} and~\ref{fig:bfs-edges} (perhaps the graph contains an additional edge, not shown, that would cause the vertex $p$ to have maximum degree), it would construct the four edge classes and four-bit labels shown in Figure~\ref{fig:bfs-edges} in its outermost call. It would then contract the labeled edges, resulting in a much smaller graph, a path of three edges shown in Figure~\ref{fig:bfs-contracted}: there are four unlabeled edges in Figure~\ref{fig:bfs-edges} but two of them form a multiple adjacency when contracted. We pass this path to the second level of recursion, which will label and contract two of the edges and leave unlabeled the third since a path has no nontrivial edge relations. In the third level of recursion, the remaining edge is labeled and contracted, leaving a single vertex in the fourth level of recursion, which terminates immediately. Thus, for this graph (which is a partial cube), the algorithm eventually terminates with seven edge classes: the four shown in Figure~\ref{fig:bfs-edges}, one for the two unlabeled edges that are part of a four-cycle in that figure, and one each for the two remaining edges.

\begin{lemma}
\label{lem:label-alg}
The algorithm above terminates in time $O(n^2)$, and either produces a partition of the edges into classes and a bitvector labeling of the vertices or terminates with the claim that $G$ is not a partial cube.
If $G$ is a partial cube, the algorithm produces a correct partition and a correct labeling of $G$.
If $G$ is not a partial cube, but the algorithm nevertheless returns a partition and a bitvector labeling, then each edge set in the partition forms a cut in the graph separating the vertices for which the bit corresponding to that edge set is 0 from the vertices for which the bit is~1.
\end{lemma}

\begin{proof}
As is standard in graph algorithms, removing self-loops and multiple adjacencies from the contracted graph $G'$ may be performed in time $O(m)$ by assigning index numbers to the vertices and then applying two rounds of bucket sorting to the list of edges, one for each endpoint of each edge. The other steps of the algorithm, except for applying Lemma~\ref{lem:some-classes} and concatenating vertex  labels, take time $O(m)$. By Lemma~\ref{lem:some-classes}, the time to find ${\mathcal E}$ is $O(dn)$, where $d$ is the number of equivalence classes found. And, the time spent in the final step of the algorithm concatenating vertex labels is also $O(dn)$.
Thus, in each recursive call of the algorithm, the time taken at that level of the recursion is $O(dn+m)=O(dn)$. Since we limit the algorithm to produce a total of at most $n-1$ classes, the total time summed over all recursive calls is at most $O(n^2)$.

If the input is a partial cube, we prove by induction on the number of recursive calls that the output is correct. As a base case, this is clearly true for the single-vertex graph. Otherwise, each call to the algorithm of Lemma~\ref{lem:some-classes} finds a valid set of classes $[pq]$, which by Lemma~\ref{lem:winkler} are equivalence classes of $\sim_G$, and a valid vertex labeling for the semicubes derived from those classes. The induction hypothesis tells us that the algorithm finds a correct labeling and partitioning for the contracted graph $G'$, and by Lemma~\ref{lem:contraction-pcube} it is also correct when translated to the corresponding objects of $G$. The algorithm simply combines these two components of a correct labeling and therefore all equivalence classes it outputs are correct. By the induction hypothesis again, every edge of $G'$ is part of one of the output equivalence classes, from which it follows that these classes when translated to $G$ include all edges not already part of a class in ${\mathcal E}$; therefore our output list of equivalence classes is not only correct but complete, and forms a partition of the edges of $G$.

If the input is not a partial cube, the desired edge cut property nevertheless follows for the edge classes in ${\mathcal E}$ by Lemma~\ref{lem:class-cut}, and can be shown to hold for all edge classes straightforwardly by induction on the number of recursive calls.
\end{proof}

\section{All pairs shortest paths}
\label{sec:apsp}

In order to verify that the given graph is a partial cube, we check that the labeling constructed by Lemma~\ref{lem:label-alg} is a correct partial cube labeling of the graph. To do this, we need distance information about the graph, which (if it is a correctly labeled partial cube) can be gathered by the all-pairs shortest paths algorithm for partial cubes from our previous paper~\cite{cs.DS/0206033}. However, as part of our verification algorithm, we will need to apply this algorithm to graphs that may or may not be partial cubes. So, both for the purpose of providing a self-contained explanation and in order to examine what the algorithm does when given an input that may not be a partial cube, we explain it again in some detail here.

It will be convenient to use some of the language of \emph{media theory}~\cite{EppFalOvc-08,FalOvc-DAM-02}, a framework for describing systems of states and actions on those states (called \emph{media}) as finite state machines satisfying certain axioms. The states and adjacent pairs of states in a medium form the vertices and edges of a partial cube, and conversely any partial cube can be used to form a medium. We do not describe here the axioms of media theory, but only borrow sufficient of its terminology to make sense of the all-pairs shortest path algorithm.

Thus, we define a \emph{token} to be an ordered pair of complementary semicubes $(S_{pq},S_{qp})$. If $G$ is a graph, with vertices labeled by bitvectors, we may specify a token as a pair $(i,b)$ where $i$ is the index of one of the coordinates of the bitvectors, $S_{pq}$ is the semicube of vertices with $i$th coordinate equal to~$b$, and $S_{qp}$ is the semicube of vertices with $i$th coordinate unequal to~$b$. A token \emph{acts} on a vertex $v$ if $v$ belongs to $S_{pq}$ and has a neighbor $w$ in $S_{qp}$; in that case, the result of the action is $w$. Our all-pairs shortest path algorithm begins by building a table indexed by (vertex,token) pairs, where each table cell lists the result of the action of a token $\tau$ on a vertex $v$ (or $v$ itself if $\tau$ does not act on $v$). Note that, if we are given any labeled graph that may or may not be a correctly labeled partial cube, we may still build such a table straightforwardly in time $O(n^2)$; if as part of this construction we find that a vertex $v$ has two or more neighbors in $S_{qp}$ we may immediately abort the algorithm as in this case the input cannot be a correctly labeled partial cube.

Define an \emph{oriented tree rooted at $r$} to be a subgraph of the input graph $G$, with an orientation on each edge, such that each vertex of $G$ except for $r$ has a single outgoing edge $vw$, and such that $w$ is formed by the action on $v$ of a token $(S_{pq},S_{qp})$ for which $r$ is a member of $S_{qp}$.

\begin{lemma}
\label{lem:at-least-Hamming}
Suppose we are given a graph $G$, a labeling of the vertices of $G$ by bitvectors, and a partition of the edges into classes, such that each class is the set of edges spanning the cut defined by one of the coordinates of the bitvectors. Then the graph distance between any two vertices $v$ and $w$ in $G$ is greater than or equal to the Hamming distance of the labels of $v$ and $w$.
\end{lemma}

\begin{proof}
For each bit in which the labels of $v$ and $w$ differ, the path from $v$ to $w$ must cross the corresponding cut in $G$ at least once. No two cuts can share the same path edge, as the cuts partition the edges. Therefore, any path from $v$ to $w$ must have at least as many edges as there are bit differences.
\end{proof}

\begin{lemma}
\label{lem:otree-is-sptree}
Suppose we are given a graph $G$, a labeling of the vertices of $G$ by bitvectors, and a partition of the edges into classes, such that each class is the set of edges spanning the cut defined by one of the coordinates of the bitvectors, and suppose that $T$ is an oriented tree rooted at $r$. Then $T$ is a shortest path tree for paths to $r$ in $G$, and each path from any vertex $s$ to $r$ in this tree has length equal to the 
Hamming distance between the labels of $s$ and $r$.
\end{lemma}

\begin{proof}
$T$ has no directed cycles, for traversing a cycle would cross the same cut in $G$ multiple times in alternating directions across the cut, while in $T$ any directed path can only cross a cut in the direction towards $r$. Thus, $T$ is a tree. The length of a path in $T$ from $s$ to $r$ at most equals the Hamming distance between the labels of $s$ and $r$, because by the same reasoning as above the path can only cross once the cuts separating $s$ and $r$ (for which the corresponding bits differ) and cannot cross any cut for which the corresponding bits of the labels of $s$ and $r$ agree.
By Lemma~\ref{lem:at-least-Hamming} any path must have length at least equal to the Hamming distance, so the paths in $T$ are shortest paths and have length equal to the Hamming distance.
\end{proof}

Our all-pairs shortest path algorithm traverses an Euler tour of a spanning tree of the input graph, making at most $2n-1$ steps before it visits all vertices of the graph, where each step replaces the currently visited node in the traversal by a neighboring node. As it does so, it maintains the following data structures:
\begin{itemize}
\item The current node visited by the traversal, $r$.
\item A doubly-linked ordered list $L$ of the tokens $(S_{pq},S_{qp})$ for which $r$ belongs to $S_{qp}$.
\item A pointer $p_v$ from each vertex $v\ne r$ to the first token in $L$ that acts on $v$.
\item A list $A_\tau$ for each token $\tau$ in $L$ of the vertices pointing to $\tau$.
\end{itemize}

\begin{lemma}
\label{lem:find-otree}
If the data structures described above are maintained correctly, we can construct an oriented tree rooted at $r$.
\end{lemma}

\begin{proof}
We set the directed edge out of each $v$ to be the result of the action of token $p_v$ on $v$.
\end{proof}

To update the data structure when traversing from $r$ to $r'$, we perform the following steps:
\begin{itemize}
\item Append the token $\tau=(S_{rr'},S_{r'r})$ to the end of $L$, set $p_r=\tau$, and add $r$ to $A_\tau$.
\item Let $\tau'$ be the token $(S_{r'r},S_{rr'})$; remove $r'$ from $A_{\tau'}$.
\item For each vertex $v\ne r$ in $A_{\tau'}$, search $L$ sequentially forward from $\tau'$ for the next token that acts on $v$. Replace $p_v$ with a pointer to that token and update the lists $A_i$ appropriately.
\item Remove $(S_{r'r},S_{rr'})$ from $L$.
\end{itemize}

We modify the algorithm in one small regard to handle the possibility that the input might not be a partial cube: if the search for the replacement for $p_v$ runs through all of list $L$ without finding any token that acts on $v$, we abort the algorithm and declare that the input is not a partial cube.

\begin{lemma}
\label{lem:apsp-alg}
If the input graph $G$ is a correctly labeled partial cube, the algorithm described above will correctly update the data structures at each step and find a shortest path tree rooted at each node. If the input graph is not a correctly labeled partial cube, but is a bitvector-labeled graph together with a partition of the edges into classes such that each class is the set of edges spanning the cut defined by one of the coordinates of the bitvectors, then the algorithm will abort and declare that the input is not a partial cube. In either case, the total running time is at most $O(n^2)$.
\end{lemma}

\begin{proof}
If the input is a partial cube, then, at any step of the algorithm, each vertex $v$ has a token in $L$ that acts on it, namely the token corresponding to the first edge in a shortest path from $v$ to $r$. Thus, the sequential search for a replacement for $p_v$, starting from a point in $L$ that is known to be earlier than all tokens acting on $v$, is guaranteed to find such a token. Thus, by Lemma~\ref{lem:find-otree} we have an oriented tree rooted at $r$ for each $r$, and by Lemma~\ref{lem:otree-is-sptree} this is a shortest path tree. 

Conversely, if the algorithm terminates with an oriented tree rooted at $r$ for each $r$, this gives us by Lemma~\ref{lem:otree-is-sptree} a shortest path tree in which each path length equals the Hamming distance of labels; since all graph distances equal the corresponding Hamming distances, the input is a partial cube. Thus, if the input were not a correctly-labeled partial cube, but satisfied the other conditions allowing us to apply Lemma~\ref{lem:otree-is-sptree}, the algorithm must at some point abort.

$L$ starts with at most $n-1$ items on it, and has at most $2n-1$ items added to it over the course of the algorithm. Thus, for each $v$, over the course of the algorithm, the number of steps performed by searching for a new value for $p_v$ is at most $3n-2$.  Thus, the total amount of time spent searching for updated values of $p_v$ is $O(n(3n-2))=O(n^2)$. The other steps of the algorithm are dominated by this time bound.
\end{proof}

\section{Testing correctness of the labeling}
\label{sec:verify}

We now put together the pieces of our partial cube recognition algorithm.

\begin{lemma}
\label{lem:validity}
If we are given a graph $G$, a labeling of the vertices of $G$ by bitvectors, and a partition of the edges into classes, such that each class is the set of edges spanning the cut defined by one of the coordinates of the bitvectors, then we can determine whether the given labeling is a valid partial cube labeling in time $O(n^2)$.
\end{lemma}

\begin{proof}
We apply the algorithm of Lemma~\ref{lem:apsp-alg}. By that Lemma, that algorithm either successfully finds a collection of shortest path trees in $G$, which can only happen when the input is a partial cube, or it aborts and declares that the input is not a partial cube. We use the presence or absence of this declaration as the basis for our determination of whether the given labeling is valid.
\end{proof}

\begin{theorem}
Let $G$ be an undirected graph with $n$ vertices. Then we may check whether $G$ is a partial cube, and if so construct a valid partial cube labeling for $G$, in time $O(n^2)$.
\end{theorem}

\begin{proof}
We use Lemma~\ref{lem:label-alg} to construct a partial cube labeling, and Lemma~\ref{lem:validity} to test its validity.
\end{proof}

\section{Implementation}
\label{sec:implementation}

As a proof of concept, we implemented the algorithms described in this paper as part of our open-source Python algorithm implementation library PADS, available online at \url{http://www.ics.uci.edu/~eppstein/PADS/}, replacing a previous implementation of an $O(mn)$-time algorithm.

\subsection{Implementation details}
The labeling phase of the new algorithm is in one Python module, PartialCube, and consists of approximately 66 lines of code within that module. The distance-checking phase of the algorithm is in a separate module, Medium, and consists of approximately 48 lines of code within that module. Additionally, a module performing breadth-first searches (written at the same time) and a previously-written module for testing bipartiteness of a graph (using depth-first search) were used as subroutines by the implementation.

The labeling algorithm described in this paper is recursive---it finds some labels, contracts the labeled edges, recursively labels the remaining graph, and then uncontracts it and in the process of uncontraction it extends the labels from the contracted graph to the original graph. However, some versions of Python are unsuited for algorithms involving deep recursion. Instead, we performed an iterative version of the algorithm that finds some edge equivalence classes, contracts the graph, and continues without recursing. Our implementation represents the partition of the edges into equivalence classes by a union-find data structure~\cite{Tar-JACM-75} (also previously implemented) in which each set element represents an edge of the input graph and each of the disjoint sets represented by the union-find data structure represents a set of edges that are all known to have the same label. Whenever our algorithm finds the equivalence classes of all of the edges incident to a single vertex using the algorithm of Section~\ref{sec:multiple}, it uses union operations to group those edges into a single set, and whenever it contracts those labeled edges and the contraction generates multiple adjacencies between a single pair of vertices, those multiple adjacencies are again grouped together by union operations and replaced in the contracted graph by a single representative edge. At the end of the algorithm, when the input graph has been contracted down to a single vertex, the sets of edges sharing the same label do not need to be constructed by uncontracting the graph, as they are exactly the sets represented by the union-find structure. The total time spent performing union-find operations, $O(n^2\alpha(n^2,m))=O(n^2)$, is not asymptotically larger than that for the rest of the algorithm.

Altogether, including comments, unit tests, and routines for other related tasks, but not including the other modules they refer to, both modules total 631 lines.

\subsection{Experimental tests}
In order to test how well our theoretical bounds match the actual performance of the implementation, we ran tests on a family of partial cubes generated from sets of random permutations.

Let $P=\{P_1,P_2,\dots, P_k\}$ be a set of permutations of the same $t$ items, and for each $k$-tuple of integers $X=(x_1,x_2,\dots x_k)$, $0\le x_i\le t$, let $S(X)$ be the set of items that appear in a position earlier than $x_i$ in at least one permutation $P_i$. Then the sets $S(X)$ generated in this way form an \emph{antimatroid}, and the graph that has one vertex for each such set and one edge for each two sets that differ in a single element is an example of a partial cube.
These graphs do not include all possible partial cubes; we chose them as test cases for two reasons: first because choosing $k$ permutations uniformly at random (with replacement) provides a convenient probability distribution with which to perform random testing, and second because efficient algorithms and a proof of concept implementation were available to generate these graphs from their defining permutations~\cite{Epp-LS}.

Our experimental data is presented in Table~\ref{tab:experiments}. Each row of the table shows, averaged over ten randomly chosen graphs, the number of vertices in the graph, the number of edges in the graph, the number of iterations performed in the first phase of the algorithm (in which the partial cube labeling is constructed), and the number of steps per vertex performed by the second phase of the algorithm (in which the labeling is tested). We also measured the average number of breadth-first-search passes needed within the first phase of the algorithm, limiting each pass to use bitvector operations with at most 32 bits per word, but for the size parameters we chose this number was not significantly different than the total number of iterations. The total time of the algorithm may be estimated by adding the numbers in the phase I and phase II columns and then multiplying the sum by the number of vertices.

\begin{table}[hbt]
	\centering
		\begin{tabular}{|c|c|c|c|c|c|}
		\hline
		$k$ & $t$ & $|V(G)|$ & $|E(G)|$ & phase I iterations & phase II steps / vertex  \\
		\hline
2 & 15 & 76.0 & 135.0 & 3.7 & 115.0 \\
2 & 20 & 125.8 & 229.6 & 4.6 & 193.8 \\
2 & 25 & 183.4 & 339.8 & 5.3 & 292.8 \\
2 & 30 & 238.4 & 444.8 & 6.6 & 376.4 \\
2 & 35 & 332.4 & 627.8 & 7.1 & 542.3 \\
2 & 40 & 465.6 & 889.2 & 7.9 & 759.6 \\
3 & 15 & 194.3 & 470.6 & 2.5 & 340.1 \\
3 & 20 & 473.6 & 1232.4 & 3.0 & 876.5 \\
3 & 25 & 773.4 & 2039.8 & 3.6 & 1445.5 \\
3 & 30 & 1264.9 & 3401.5 & 4.0 & 2393.8 \\
4 & 15 & 466.0 & 1410.7 & 2.1 & 879.1 \\
4 & 20 & 1192.4 & 3843.0 & 2.0 & 2306.1 \\
5 & 15 & 1029.5 & 3719.7 & 1.9 & 2008.7 \\
		\hline
		\end{tabular}
         \smallskip
	\caption{Experimental test results from our implementation, on antimatroids generated from $k$ random permutations of a $t$-item set.}
	\label{tab:experiments}
\end{table}

Because the worst case for our algorithm is a path graph, a special case of a tree, we also performed experiments for randomly generated trees (Table~\ref{tab:trees}). The trees in our experiments are rooted and ordered (that is, the ordering of the children of each node is significant for determining whether two trees are the same). The number of trees with a given number of nodes is counted by the Catalan numbers, and we generated trees uniformly at random, with probability inversely proportional to the Catalan numbers. Again, each row of the data is an average of ten trials. For these experiments, on sufficiently large trees, the phase I iteration counts diverged from the count of the number of passes over the graph that would be required using bitvectors that are limited to 32 bits. However both counts were still close to each other and both were significantly smaller than the number of vertices in the graph.

\begin{table}[hbt]
	\centering
		\begin{tabular}{|c|c|c|c|}
		\hline
		$n$ & phase I iterations & phase I 32-bit passes & phase II steps / vertex  \\
		\hline
50 & 9.6 & 9.6 & 24.5 \\
100 & 14.0 & 14.0 & 49.5 \\
200 & 21.5 & 21.5 & 99.5 \\
400 & 27.9 & 28.3 & 199.5 \\
800 & 45.5 & 49.4 & 399.5 \\
		\hline
		\end{tabular}
         \smallskip
	\caption{Test results on random $n$-node trees.}
	\label{tab:trees}
\end{table}

Our experiments showed that, for both types of graphs that we tested, the number of phase I iterations appears to be significantly less than its worst case bound of $\Theta(n)$ (a bound achieved, for instance, by path graphs); therefore, this phase of the algorithm is significantly faster than its worst case bound.
The number of steps per vertex in phase II of the algorithm was smaller for trees than it was for antimatroids, but in both test sets this number appeared to be growing linearly in the number of vertices. Therefore, for these graphs, the average performance of phase II was no better than its worst case.

\section{Conclusions}
\label{sec:conclusions}

We have shown that recognition of partial cubes may be performed in quadratic time, improving previous algorithms for the same problem. If an explicit partial cube labeling of the vertices is required as output, this is close to the $O(n^2)$ bit complexity of the output. Although not simple, our algorithms are implementable.

A worst case input, requiring $\Omega(n^2)$ bits of output and forcing the labeling phase of our algorithm to take $\Omega(n^2)$ time, takes the form of a path graph.  However, the existence of algorithms for recognizing median graphs~\cite{ImrKlaMul-SJDM-99} that do not output a labeling explicitly and are significantly faster than quadratic suggests that it may similarly be possible to recognize partial cubes more quickly. Our experiments suggest that the labeling phase of our algorithm is often much faster than our worst case bounds; however, to speed up the overall algorithm, we must also speed up the verification phase of the algorithm, which in our experimental tests was no better on average than it is in the worst case. 

\section*{Acknowledgments}
This research was supported in part by the National Science
Foundation under grant 0830403, and by the
Office of Naval Research under MURI grant N00014-08-1-1015. A preliminary version of this paper was presented in the \emph{19th ACM-SIAM Symposium on Discrete Algorithms}, 2008; some of the material from the preliminary version was also incorporated into the book \emph{Media Theory: Applied Interdisciplinary Mathematics}~\cite{EppFalOvc-08}.

\clearpage

\raggedright
\bibliographystyle{abuser}
\bibliography{media}

\begin{thebibliography}{10}

\bibitem{AloYusZwi-JACM-95}
N.~Alon, R.~Yuster, and U.~Zwick.
\newblock {Color-coding}.
\newblock {\em J. ACM} 42(4):844{--}856, 1995,
  \href{http://dx.doi.org/10.1145/210332.210337}%
{doi:10.1145/210332.210337}.

\bibitem{AurHag-MST-95}
F.~Aurenhammer and J.~Hagauer.
\newblock {Recognizing binary Hamming graphs in $O(n^2\log n)$ time}.
\newblock {\em Mathematical Systems Theory} 28:387{--}395, 1995.

\bibitem{Ava-PAMS-61}
S.~P. Avann.
\newblock {Metric ternary distributive semi-lattices}.
\newblock {\em Proc. Amer. Math. Soc.} 12(3):407{--}414, 1961,
  \href{http://dx.doi.org/10.2307/2034206}%
{doi:10.2307/2034206}.

\bibitem{BanCheEpp-SJDM-10}
H.-J. Bandelt, V.~Chepoi, and D.~Eppstein.
\newblock {Combinatorics and geometry of finite and infinite squaregraphs}.
\newblock {\em SIAM J. Discrete Math.} 24(4):1399{--}1440, 2010,
  \href{http://dx.doi.org/10.1137/090760301}%
{doi:10.1137/090760301},
  \href{http://arxiv.org/abs/0905.4537}{arXiv:0905.4537}.

\bibitem{BanCheEpp-10}
H.-J. Bandelt, V.~Chepoi, and D.~Eppstein.
\newblock {Ramified rectilinear polygons: coordinatization by dendrons}.
\newblock ACM Computing Research Repository, 2010,
  \href{http://arxiv.org/abs/1005.1721}{arXiv:1005.1721}.

\bibitem{BanMacRic-MPE-00}
H.-J. Bandelt, V.~Macaulay, and M.~B. Richards.
\newblock {Median networks: speedy construction and greedy reduction, one
  simulation, and two case studies from human mtDNA}.
\newblock {\em Molecular Phylogenetics and Evolution} 16(1):8{--}28, 2000,
  \href{http://dx.doi.org/10.1006/mpev.2000.0792}%
{doi:10.1006/mpev.2000.0792}.

\bibitem{BanVel-PLMS-89}
H.-J. Bandelt and M.~van~de Vel.
\newblock {Embedding topological median algebras in product of dendrons}.
\newblock {\em Proc. London Math. Soc.} 58(3):439{--}453, 1989.

\bibitem{BenFar-LATIN-00}
M.~A. Bender and M.~Farach-Colton.
\newblock {The LCA problem revisited}.
\newblock {\em Proc. 4th Latin American Symp. on Theoretical Informatics},
  pp.~88{--}94. Springer-Verlag, Lecture Notes in Computer Science 1776, 2000,
  \href{http://dx.doi.org/10.1007/10719839\_9}%
{doi:10.1007/10719839\_9}.

\bibitem{BirKis-BAMS-47}
G.~Birkhoff and S.~A. Kiss.
\newblock {A ternary operation in distributive lattices}.
\newblock {\em Bull. Amer. Math. Soc.} 52(1):749{--}752, 1947,
  \href{http://dx.doi.org/10.1090/S0002-9904-1947-08864-9}%
{doi:10.1090/S0002-9904-1947-08864-9}.

\bibitem{BreImrKla-DAM-03}
B.~Bre{\v{s}}ar, W.~Imrich, and S.~Klav{\v{z}}ar.
\newblock {Fast recognition algorithms for classes of partial cubes}.
\newblock {\em Discrete Applied Mathematics} 131(1):51{--}61, 2003,
  \href{http://dx.doi.org/10.1016/S0166-218X(02)00416-X}%
{doi:10.1016/S0166-218X(02)00416-X}.

\bibitem{Bun-MAHS-71}
P.~Buneman.
\newblock {The recovery of trees from measures of dissimilarity}.
\newblock {\em Mathematics in the Archaeological and Historical Sciences},
  pp.~387{--}395. Edinburgh University Press, 1971.

\bibitem{Djo-JCTB-73}
D.~Z. Djokovic.
\newblock {Distance preserving subgraphs of hypercubes}.
\newblock {\em J. Combinatorial Theory, Ser. B} 14:263{--}267, 1973.

\bibitem{DoiFal-99}
J.-P. Doignon and J.-C. Falmagne.
\newblock {\em {Knowledge Spaces}}.
\newblock Springer-Verlag, 1999.

\bibitem{Epp-GD-04}
D.~Eppstein.
\newblock {Algorithms for drawing media}.
\newblock {\em Proc. 12th Int. Symp. Graph Drawing (GD 2004)}, pp.~173{--}183.
  Springer-Verlag, Lecture Notes in Computer Science 3383, 2004,
  \href{http://arxiv.org/abs/cs.DS/0406020}{arXiv:cs.DS/0406020}.

\bibitem{Epp-EJC-05}
D.~Eppstein.
\newblock {The lattice dimension of a graph}.
\newblock {\em Eur. J. Combinatorics} 26(5):585{--}592, 2005,
  \href{http://dx.doi.org/10.1016/j.ejc.2004.05.001}%
{doi:10.1016/j.ejc.2004.05.001},
  \href{http://arxiv.org/abs/cs.DS/0402028}{arXiv:cs.DS/0402028}.

\bibitem{Epp-GD-08}
D.~Eppstein.
\newblock {Isometric diamond subgraphs}.
\newblock {\em Proc. 16th Int. Symp. Graph Drawing}, pp.~384{--}389.
  Springer-Verlag, Lecture Notes in Computer Science 5417, 2008,
  \href{http://dx.doi.org/10.1007/978-3-642-00219-9\_37}%
{doi:10.1007/978-3-642-00219-9\_37},
  \href{http://arxiv.org/abs/0807.2218}{arXiv:0807.2218}.

\bibitem{Epp-LS}
D.~Eppstein.
\newblock {Learning sequences}.
\newblock ACM Computing Research Repository, 2008,
  \href{http://arxiv.org/abs/0803.4030}{arXiv:0803.4030}.

\bibitem{Epp-JGAA-08}
D.~Eppstein.
\newblock {Upright-quad drawing of $st$-planar learning spaces}.
\newblock {\em J. Graph Algorithms {\&} Applications} 12(1):51{--}72, 2008,
  \href{http://arxiv.org/abs/cs.CG/0607094}{arXiv:cs.CG/0607094},
  \url{http://www.jgaa.info/accepted/2008/Eppstein2008.12.1.pdf}.

\bibitem{Epp-JCG-10}
D.~Eppstein.
\newblock {Happy endings for flip graphs}.
\newblock {\em Journal of Computational Geometry} 1(1):3{--}28, 2010,
  \href{http://arxiv.org/abs/cs.CG/0610092}{arXiv:cs.CG/0610092}.

\bibitem{cs.DS/0206033}
D.~Eppstein and J.-C. Falmagne.
\newblock {Algorithms for media}.
\newblock ACM Computing Research Repository, 2002,
  \href{http://arxiv.org/abs/cs.DS/0206033}{arXiv:cs.DS/0206033}.

\bibitem{EppFalOvc-08}
D.~Eppstein, J.-C. Falmagne, and S.~Ovchinnikov.
\newblock {\em {Media Theory: Applied Interdisciplinary Mathematics}}.
\newblock Springer-Verlag, 2008.

\bibitem{EppWor-09}
D.~Eppstein and K.~A. Wortman.
\newblock {Optimal angular resolution for face-symmetric drawings}.
\newblock ACM Computing Research Repository, 2009,
  \href{http://arxiv.org/abs/0907.5474}{arXiv:0907.5474}.

\bibitem{FalOvc-DAM-02}
J.-C. Falmagne and S.~Ovchinnikov.
\newblock {Media theory}.
\newblock {\em Discrete Applied Mathematics} 121(1{--}3):103{--}118, 2002.

\bibitem{Fed-MotAMS-95}
T.~Feder.
\newblock {\em {Stable Networks and Product Graphs}}.
\newblock Memoirs of the American Mathematical Society 555. 1995.

\bibitem{GavPelPer-Algs-04}
C.~Gavoille, D.~Peleg, S.~P{\'e}rennes, and R.~Raz.
\newblock {Distance labeling in graphs}.
\newblock {\em J. Algorithms} 53(1):85{--}112, 2004,
  \href{http://dx.doi.org/10.1016/j.jalgor.2004.05.002}%
{doi:10.1016/j.jalgor.2004.05.002}.

\bibitem{GraPol-BSTJ-71}
R.~L. Graham and H.~Pollak.
\newblock {On addressing problem for loop switching}.
\newblock {\em Bell System Technical Journal} 50:2495{--}2519, 1971.

\bibitem{HagImrKla-TCS-99}
J.~Hagauer, W.~Imrich, and S.~Klav{\v{z}}ar.
\newblock {Recognizing median graphs in subquadratic time}.
\newblock {\em Theoret. Comput. Sci.} 215:123{--}136, 1999.

\bibitem{HarTar-SJC-84}
D.~Harel and R.~E. Tarjan.
\newblock {Fast algorithms for finding nearest common ancestors}.
\newblock {\em SIAM J. Comput.} 13(2):338{--}355, 1984,
  \href{http://dx.doi.org/10.1137/0213024}%
{doi:10.1137/0213024}.

\bibitem{ImrKla-BICA-93}
W.~Imrich and S.~Klav{\v{z}}ar.
\newblock {A simple $O(mn)$ algorithm for recognizing Hamming graphs}.
\newblock {\em Bull. Inst. Combin. Appl.} 9:45{--}56, 1993.

\bibitem{ImrKlaMul-SJDM-99}
W.~Imrich, S.~Klav{\v{z}}ar, and H.~M. Mulder.
\newblock {Median graphs and triangle-free graphs}.
\newblock {\em SIAM J. Discrete Math.} 12:111{--}118, 1999.

\bibitem{KlaGutMoh-JoCIaCS-95}
S.~Klav{\v{z}}ar, I.~Gutman, and B.~Mohar.
\newblock {Labeling of benzenoid systems which reflects the vertex-distance
  relations}.
\newblock {\em Journal of Chemical Information and Computer Sciences}
  35(3):590{--}593, 1995.

\bibitem{Mat-Comb-06}
J.~Matou{\v{s}}ek.
\newblock {The number of unique-sink orientations of the hypercube}.
\newblock {\em Combinatorica} 26(1):91{--}99, 2006,
  \href{http://dx.doi.org/10.1007/s00493-006-0007-0}%
{doi:10.1007/s00493-006-0007-0}.

\bibitem{Neb-CMUC-71}
L.~Nebesk'y.
\newblock {Median graphs}.
\newblock {\em Comment. Math. Univ. Carolinae} 12:317{--}325, 1971.

\bibitem{math.CO/0402246}
S.~Ovchinnikov.
\newblock {The lattice dimension of a tree}.
\newblock Electronic preprint arxiv:math.CO/0402246, 2004.

\bibitem{Ovc-JMP-05}
S.~Ovchinnikov.
\newblock {Hyperplane arrangements in preference modeling}.
\newblock {\em J. Mathematical Psychology} 49:481{--}488, 2005.

\bibitem{PolRou-ICMC-76}
O.~E. Polansky and D.~H. Rouvray.
\newblock {Graph-theoretical treatment of aromatic hydrocarbons. II. The
  analysis of all-benzenoid systems}.
\newblock {\em Informal Communications in Mathematical Chemistry, Match No. 2},
  pp.~111{--}115. Inst. f{\"u}r Strahlenchemie, Max-Planck-Institut f{\"u}r
  Kohlenforschung, M{\"u}lhein a.d. Ruhr, 1976.

\bibitem{Tar-JACM-75}
R.~E. Tarjan.
\newblock {Efficiency of a good but not linear set union algorithm}.
\newblock {\em J. ACM} 22(2):215{--}225, 1975,
  \href{http://dx.doi.org/10.1145/321879.321884}%
{doi:10.1145/321879.321884}.

\bibitem{Win-DAM-84}
P.~Winkler.
\newblock {Isometric embeddings in products of complete graphs}.
\newblock {\em Discrete Applied Mathematics} 7:221{--}225, 1984.

\end{thebibliography}

\end{document}